# On the origin of non-Arrhenius behavior of grain growth

Xinlei Pan[1], Jingyu Li[1], Jianfeng Hu[1]*

[1] School of Materials Science and Engineering, Shanghai University; Shanghai 200444, China.

*Correspondence Email: jianfenghu@shu.edu.cn

**Abstract**

Non-Arrhenius grain growth has been observed in a range of polycrystalline materials; however, its fundamental mechanisms—particularly whether the process is thermally activated or exhibits anti-thermally activation—remain controversial. In this study, $SrTiO_3$ was employed as a model system to systematically investigate non-Arrhenius grain growth behavior through combined experimental and theoretical approaches, utilizing a newly developed grain growth model. The results reveal that non-Arrhenius grain growth is a thermally activated process without a definitive characteristic temperature, which is primarily controlled by the interplay between temperature-dependent factors and the temperature-independent parameters such as grain size and its distribution. Moreover, during abnormal grain growth (AGG), the non-Arrhenius behavior of grain growth primarily occurs at lower temperatures and gradually transitions to Arrhenius-type behavior as the temperature increases.

## 1. Introduction

Grain growth behavior plays a critical role in microstructure evolution and grain size control of polycrystalline materials. Over past decades, extensive research has advanced the understanding of grain growth [1], leading to the development of various theoretical models such as Hillert's theory [2] and von Neumann-Mullins equation [3]. These classical grain growth models are all based on the grain boundary (GB) migration formula in reaction rate theory as follows [4],

$$v = Mp = M_0 \exp\left(-\frac{Q}{k_B T}\right) \cdot p \qquad 1)$$

where $M$ is GB mobility, which is the coefficient between GB migration velocity $v$ and the driving force $p$. $M_0$ is a pre-exponential constant. $Q$ is an intrinsic barrier (or activation energy) for GB migration. $T$ and $k_B$ are the temperature and the Boltzmann constant, respectively. However, these classical formulas can only explain the normal grain growth with mono-modal grain size distribution (GSD) but cannot adequately describe AGG with bimodal distributions [5–7]. Recently, Hu and coworkers proposed a new GB migration formula, based on which a grain growth formula was established[5,8]. It effectively explains the general grain growth behaviors, including normal, abnormal, and stagnant grain growth. According to this model, the classical Eq. 1 is only valid under specific conditions above GB roughening transition temperature.

In current grain growth formulas, GB mobility, the coefficient linking migration velocity and driving force, follows a thermally activated Arrhenius relationship. Thus, grain growth is widely accepted to exhibit Arrhenius-type behavior, with higher annealing temperatures yielding larger grain sizes, as showed by extensive experimental observations. However, in recent years, counter-intuitive grain growth (or coarsening) behavior has been observed in grain growth experiments across various materials[9–12]. Specifically, within certain temperature ranges, specimens sintered at higher temperatures exhibited smaller grain sizes compared to those processed at lower temperatures, a phenomenon termed as non-Arrhenius grain growth. This differs from the behaviors of normal and abnormal grain growth, which are distinguished by a monomodal or bimodal grain size distribution. For instance, anomalous non-Arrhenius grain growth has been observed in SrTiO$_3$ ceramic within the annealing temperature range of about 1350 °C-1425 °C [9,10,13,14]. Similar phenomena have been documented in (K, Na)NbO$_3$-based ceramics [11]. Additionally, in nanocrystalline copper, specimens annealed at very low temperature (-190 °C) exhibited larger grain sizes compared to those processed at room temperature [12]. Nevertheless, this counter-intuitive non-Arrhenius grain growth has not yet been adequately understood.

Several mechanisms have been proposed to explain this counter-intuitive non-Arrhenius grain growth behavior. For $SrTiO_3$ ceramics, the non-Arrhenius behavior has been attributed to co-existing fast and slow GB types with different mobilities [10,15]. This explanation postulates that GB migration exhibits anti-Arrhenius (or anti-thermal) behavior within the temperature range of 1350 °C-1425 °C, contrary to the relationship described by Eq. 1. Specifically, GB mobility exhibits an anti-Arrhenius characteristic, resulting in the migration velocities of most GBs at lower temperatures that exceed those observed at higher temperatures. However, this model does not adequately address the structural and chemical characteristics that distinguish fast and slow GBs, nor does it elucidate the conditions and mechanisms governing their transformation with temperature. While studies have indicated differences in GB energies between large and small grains in $SrTiO_3$ specimens, these differences are not significant [13]. Additionally, detailed high-resolution transmission electron microscopy (HRTEM) investigations have revealed no substantial differences in characteristics between GBs above and below the transition temperature [16]. Furthermore, according to this explanation, one must also address the challenge of explaining how a growing grain can consistently maintain its fast GB or slow GB character throughout its growth, especially given that the grain will inevitably encounter a wide variety of GB types during the growth process.

$$v = v_0 \exp\left(-\frac{Q}{k_B T}\right) 2\sinh\left(\frac{p}{2k_B T}\right) \qquad 2)$$

Recently, researchers utilizing Eq. 2, which shares a simplified form with Eq. 1, found that when the activation energy $Q$ for GB migration is very small, non-Arrhenius GB migration behavior can inherently occur without assuming the coexistence of different GB types [17]. Specifically, GB migration velocity at higher temperatures can be lower than at lower temperatures. In this case, the mobility described in Eq. 2 remains a thermally activated parameter, which fundamentally differs from the assumption adopted in the aforementioned model. However, the activation energy $Q$ required for non-Arrhenius GB migration behavior, as suggested by Eq. 2, is significantly smaller than the experimentally measured activation energies for GB migration [4]. Moreover, the calculated transition temperature for non-Arrhenius GB migration (no more than 230 °C) is much lower than the transition temperature observed in ceramics exhibiting non-Arrhenius grain growth behavior. Meanwhile, the high driving force required for a pronounced transition in GB migration velocities (comparable to the capillary force in nanocrystalline materials) is noticeably higher than the curvature-driven forces typically associated with grain growth during high-temperature sintering [18]. Furthermore, Eq. 2 fails to explain the phenomenon of grain growth returning to an Arrhenius-type behavior at even higher temperatures following the transition to non-Arrhenius behavior. These limitations suggest that Eq. 2 is not suitable for explaining commonly observed non-Arrhenius grain growth in ceramics.

Furthermore, experimental observations revealed that after the occurrence of non-Arrhenius grain growth behavior, Arrhenius-type grain growth behavior reemerged as the temperature increases [10,14]. Notably, both types of grain growth behavior occurred during AGG. The aforementioned mechanisms fail to explain why non-Arrhenius grain growth occurs during the AGG and do not address the intrinsic correlation between non-Arrhenius grain growth and AGG. Furthermore, these two mechanisms adopt contradictory perspectives regarding whether the mobility exhibits anti-Arrhenius behavior. Therefore, this study aims to investigate the underlying mechanisms of non-Arrhenius grain growth behavior and its intrinsic relationship with AGG. A detailed examination of the onset and termination processes of non-Arrhenius grain growth behavior is conducted using $SrTiO_3$ ceramics as a model system. Furthermore, the recently developed grain growth model is utilized for the first time to investigate the physical mechanisms underlying non-Arrhenius behavior. This work provides theoretical insights for microstructural design utilizing the non-Arrhenius phenomenon.

## 2. Experimental and Methods

### 2.1 Experimental

The starting powder of SrTiO$_3$ powder was commercially available with an average size of 34 nm (Nanoxide TM HPS 1000, TPL, Inc., USA). The sintering processes were carried out in a sparking plasma sintering (SPS) apparatus, LABOX-1575 (Sinter Land Inc., Japan). A pressure of 75 MPa was applied at the beginning of the sintering cycles. The samples were automatically raised to 600 °C, and then were directly heated up to the sintering temperatures with the heating rates of 10 °C/ min which monitored and regulated by an optical pyrometer focused on the surface of the die. The holding time was 0, 5, 15, 30 or 120 min at sintering temperatures that followed by natural cooling to room temperature. This SPS process allows a cooling rate over 320 °C/min to retain high-temperature microstructure. All samples were annealed in air at 700 °C for 2 hours to remove carbon contamination. This annealing temperature is safe to avoid any grain growth. Both polished and fracture surfaces of the prepared samples were examined in scanning electron microscopes (SEM-FEG Gemini 300, Carl Zeiss). The grain sizes in the polycrystalline samples were conventionally measured using the line intersection method on SEM micrographs with the same resolution (at least 250 grains for each sample).

### 2.2 Methods and simulations

The simulations are all based on the recently-developed grain growth equation as follows [5,8]

$$\frac{dD}{dt} = M\gamma(n-1)\frac{1}{D} \cdot e^{\left(-C \cdot \frac{\varepsilon^* D}{(n-1)T}\right)} \qquad 3)$$

Here, $D$ is the linear dimension of grain, while $n$ is the ratio of the curvatures on either side of a grain boundary, serving as a dimensionless parameter reflecting GSD characteristics in polycrystalline materials. $\varepsilon^*$ denotes the effective GB step energy. $C$ and $T$ are the constant coefficient and the absolute temperature, respectively. To calculate the evolution of grain size $D$ as a function of annealing time $t$, the above Eq. 3 is integrated to yield the following expression:

$$e^{C_1 \cdot D} \cdot (C_1 \cdot D - 1) = e^{C_1 \cdot D_0} \cdot (C_1 \cdot D_0 - 1) + C_1^2 \cdot C_2 \cdot t \qquad 4)$$

In this equation, $D_0$ is the initial linear size of a grain, and $t$ is the annealing time. The coefficients $C_1$ and $C_2$ are defined as $C_1 = \left(\frac{C \cdot \varepsilon^*}{T(n-1)}\right)$ and $C_2 = M\gamma(n-1)$, respectively. This equation is a type of Lambert W function that is commonly encountered in engineering, physics, and statistics. The Eq. 4 enables an accurate determination of grain growth over time. The equation provides a precise description of the variation in grain size with annealing time and can be used to calculate the grain size at different annealing durations. Furthermore, based on Eq. 4, a program was developed using the Matlab software to numerically simulate the grain growth process. The typical values of constant coefficient were used in these numerical simulations, such as $M_0 = 1.51 \times 10^{-6}$ m$^4$/(J.s), $Q = 70$ kJ/mol and $\gamma = 0.1$ J/m$^2$. In these simulations, the time increment for grain growth calculations was set to 12 seconds.

### 3. Results and discussions

#### 3.1 Experimental observation of grain growth behavior

The grain growth behavior of nano-grain SrTiO$_3$ at different sintering temperatures ranging from 925 °C to 1000 °C is illustrated in Fig. 1. The results demonstrate that AGG occurred in samples at all sintering temperatures during the 15-minute holding period. Samples sintered at 925 °C and 950 °C exhibit only a small number of abnormally grown grains within the 15-minute holding time.

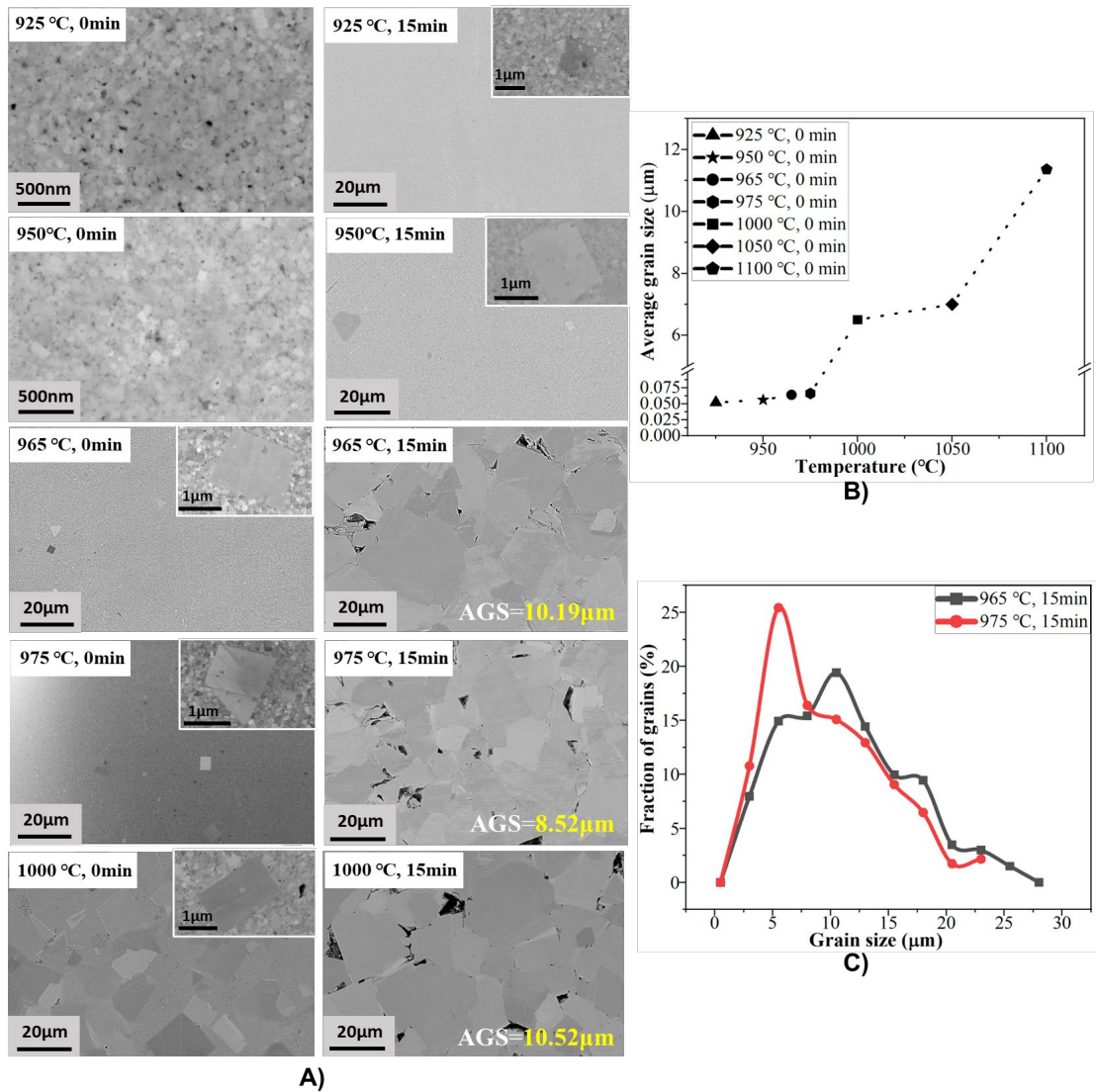

Fig. 1. Grain growth behavior in SrTiO$_3$ ceramics. **A)** SEM backscattered electron micrographs of sintered samples under different sintering temperatures and holding times. AGS denotes average grain size. **B**) Temperature dependence curve of AGS for samples at a holding time of 0 minutes. **C**) Comparative grain size distribution in samples sintered at 965 °C and 975 °C with 15-minute holding time.

Comparatively, the 950 °C sample shows slightly more and larger abnormally grown grains, as shown in Fig. 1A. For abnormally growing grains without impingement each other, their size increased rapidly with higher sintering temperatures or longer holding times. As the sintering temperature further increased, grain growth at higher sintering temperatures notably accelerated, accompanied by an increase in the number of growing grains. At a holding time of zero minutes, a small number of large, abnormally grown grains (already exceeding the micron scale) emerged in the 965 °C and 975 °C samples, comparable to those observed in samples sintered at 950 °C for 15 minutes. For samples sintered at temperatures of 1000 °C and above, a significant number of abnormally grown grains were observed, most exceeding 5 microns in size and impinging on one another. Within the 15-minute holding period, the GSD of samples sintered at 965 °C and above changed from bi-modal to mono-modal distribution, with matrix grains being almost entirely consumed. These results indicate that during the initial few minutes of holding, the average grain size of abnormally growing grains in these samples accelerated growth with increasing temperature, exhibiting Arrhenius-type grain growth behavior as shown in Fig. 1B. However, as the holding time extended to 15 minutes, a counter-intuitive non-Arrhenius grain growth behavior was observed in the samples sintered at 965 °C and 975 °C. Specifically, the grain size of the sample sintered at 975 °C was smaller than that of the sample sintered at 965 °C, as shown in Fig. 1C. This unexpected phenomenon occurs at a temperature significantly lower than the reported 1390 °C, at which $SrTiO_3$ ceramics exhibit non-Arrhenius grain growth behavior in the literature[15].

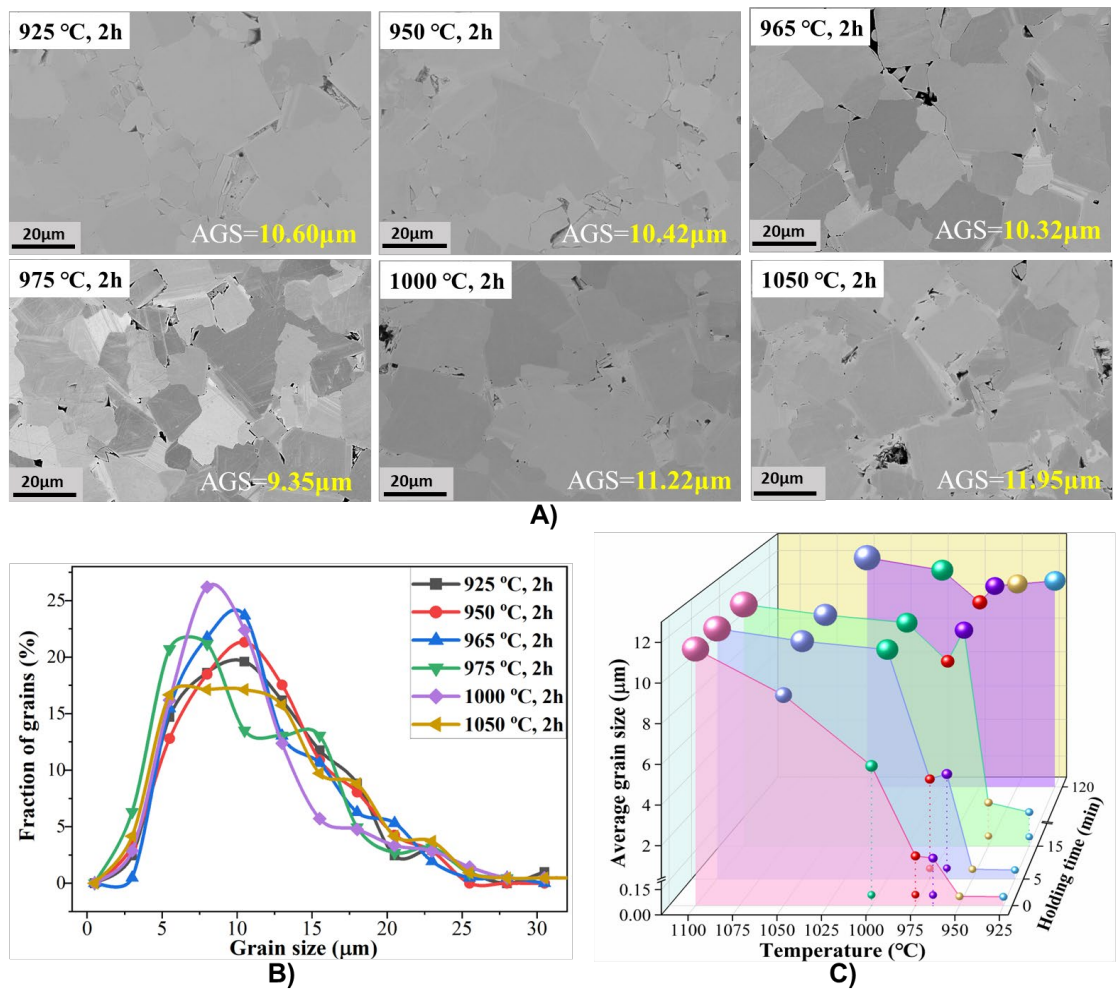

Fig. 2. **A**) SEM backscattered electron micrographs of samples held for 2 hours at different sintering temperatures. **B**) Grain size distribution of samples sintered at various temperatures. **C**) An overview of AGS in samples with different sintering temperatures and holding times.

When the holding time was further extended to two hours, all sintered samples (especially those sintered at 925 °C and 950 °C) transitioned to a mono-modal GSD, and the grain size of all samples increased, as shown in Fig. 2A-2B. However, apart from the previously bi-modal distributed samples sintered at 925 °C and 950 °C, the increase in grain size for other sintered samples is minimal. Further comparison reveals that the grain size of the sample sintered at 975 °C remained the smallest among these samples in Fig. 2C. Surprisingly, for samples sintered below 975 °C, a lower sintering temperature resulted in larger final grain sizes, as shown in Fig. 2. This indicates that the onset temperature for non-Arrhenius grain growth behavior decreases correspondingly. Notably, as described previously, the abnormally growing grains in these samples displayed Arrhenius-type growth behavior during the initial few minutes of dwell time prior to mutual impingement. In contrast, for samples sintered at temperatures above 975 °C, the grain size increased with temperature and continued to follow Arrhenius-type grain growth behavior. These findings indicate that the occurrence of non-Arrhenius grain growth behavior does not correspond to a specific temperature, implying that non-Arrhenius grain growth behavior is not solely determined by temperature-dependent variables.

**3.2 Simulation of grain growth behavior**

As shown in Fig. 2C, the non-Arrhenius grain growth behavior occurs during the AGG stage, suggesting a strong correlation with AGG. Since Eq. 3 is capable of explaining AGG [5,8], this formula should also be applicable in explaining non-Arrhenius grain growth behavior. The initial stage of grain growth, during which abnormally growing grains have not yet impinged on one another, was simulated using Eq. 4 for an identical polycrystalline system at different temperatures (corresponding to different GB step energies). As shown in Fig. 3, the simulation results reveal that the inherent coexistence of growing and stagnant grains in the polycrystalline system when the GB step energy is nonzero. The growing grains predominantly appear in regions with larger grain sizes, indicating that AGG occurs across these temperature ranges. However, noticeable differences exist among them. At lower temperatures, only a small proportion of larger grains can achieve preferential growth, requiring an initial incubation period before displaying slow growth at a low rate, as shown in Fig. 3A. As the temperature increases, the region of growing grains extends noticeably to include smaller grains (*i.e.*, the fraction of growing grains increases significantly), while the grain growth rate simultaneously accelerates. Moreover, the incubation period for grain growth is significantly shortened with increasing temperature, as shown in Fig. 3B-3D. With prolonged holding time, the abnormally growing grains continued to accelerate growth by consuming the stagnant matrix grains. At this stage, grain growth exhibits Arrhenius-type behavior, with grain size increasing as the temperature rises. These simulation results align with the observations made from the samples soaked during the initial few minutes of the experiments shown in Fig. 1B.

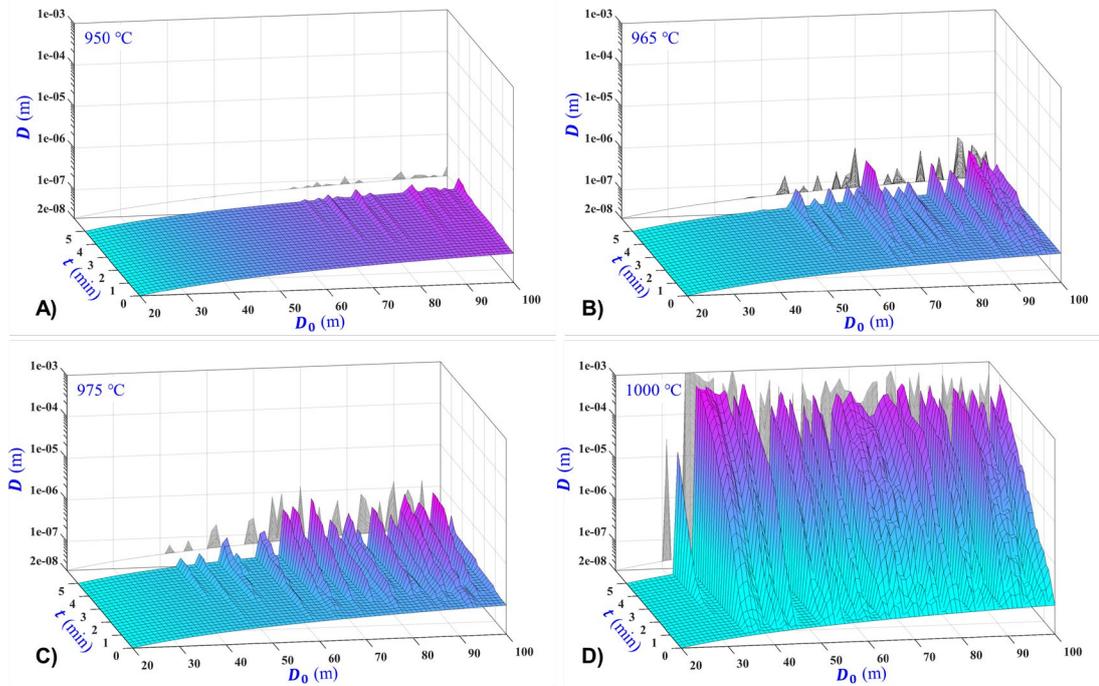

Fig. 3. Simulation of initial-stage grain growth behavior in polycrystalline systems at different temperatures. **A)** 950 °C, **B)** 965 °C, **C)** 975 °C and **D)** 1000 °C.

When most of the matrix grains were consumed, the growing grains began to impinge with each other, and the GSD transitioned from a bi-modal to a mono-modal distribution. At this stage, the size ratio between neighboring grown grains in the polycrystalline system decreases significantly (*i.e.*, the parameter $n$ in Eq. 3 decreases significantly according to its definition). To investigate the growth behavior of abnormally grown grains following mutual impingement, growth simulations were conducted by introducing abnormally grown grains of varying sizes into a polycrystalline system with a mono-modal GSD, assuming the value of $n$ for abnormally grown grains to be less than or equal to 5, at the same temperature as shown in Fig. 4.

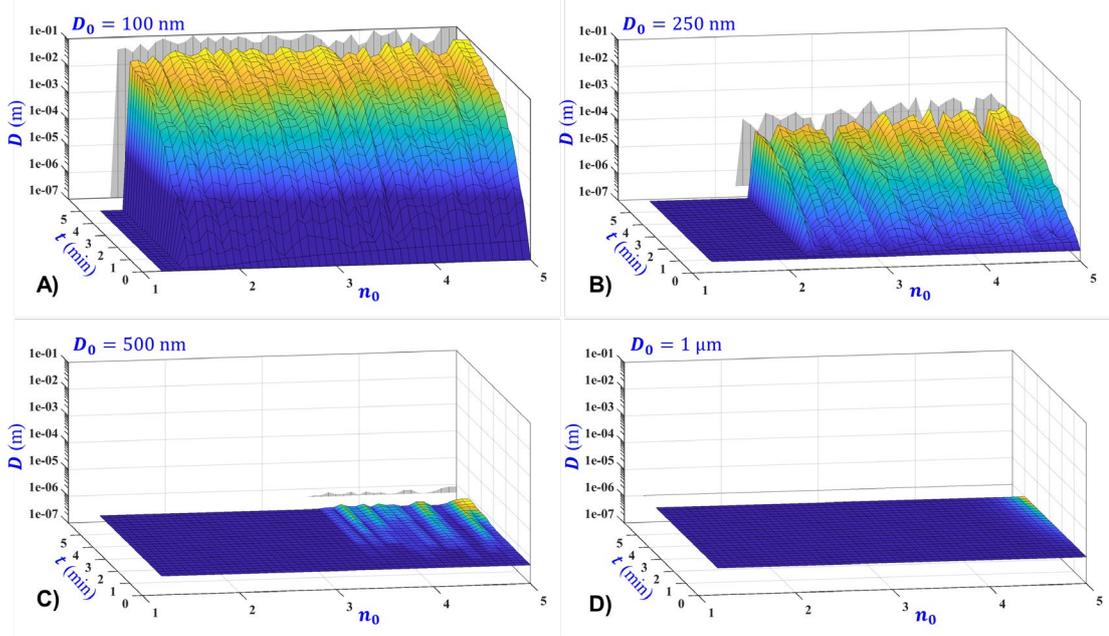

Fig. 4. Numerical simulation of the growth behavior of individual grains with different initial sizes ($D_0$) in a polycrystalline system under the same temperature and $n$-value range conditions. **A)** 100 nm, **B)** 250 nm, **C)** 500 nm and **D)** 1μm. This simulation reveals how grain size influences the growth behavior of abnormally growing grains when the $n$-value decreases sharply following mutual impingement.

When the size of growing grain is very small, a low value of $n$ is sufficient to sustain its growth. Thus, even after mutual impingement reduces the $n$ value, grains of this size can usually accelerate their growth further by consuming smaller, grown neighboring grains, as shown in Fig. 4A. As the size of the growing grain increases, the threshold value of $n$ required for its continued growth also increases, and the incubation period becomes correspondingly longer as shown in Fig. 4B-4C. Consequently, the number fraction of grown grains able to continue growing after mutual impingement is significantly reduced, and the incubation period becomes longer. When the grain size increases to a certain size and the matrix grains are completely consumed, all abnormally grown grains stagnate in growth after impingement each other, leading to a stagnation state with a mono-modal distribution in Fig. 4D. This result aligns with the experimental observations after a 2-hour holding period in Fig. 2. Therefore, the simulation results demonstrate that the grain growth behavior observed in the above experiments can be effectively explained using Eq. 3. These results further reveal that, as temperature increases, the main changes are a significant increase in the fraction of abnormally growing grains and a notable shortening of the incubation period. Since non-Arrhenius grain growth occurs during the AGG stage, its emergence is correlated to the number fraction of abnormally growing grains and their incubation period.

**3.3 Origin of non-Arrhenius grain growth behavior**

The simulation results above reveal that the occurrence of non-Arrhenius grain growth is related to the number fraction of abnormally growing grains. This findings aligns with the phase-field simulation results reported in previous literature [10], where non-Arrhenius grain growth behavior was attributed to the coexistence of varying fractions of fast-growing grains with high GB mobility and slow-growing grains with low GB mobility. However, the simulations based on Eq. 4 demonstrate that, in polycrystalline systems where the effective GB step energy $\varepsilon^*$ is non-zero, the intrinsic coexistence of growing and stagnant grains can arise even when GB energy and mobility are uniform throughout the system. This indicates that the coexistence of distinct GB types (or mobilities) in polycrystalline system is not required for the emergence of non-Arrhenius grain growth behavior. The changes in temperature (or GB step energy) alone can trigger the emergence or disappearance of non-Arrhenius grain growth behavior as shown in Fig. 2C. According to Eq. 3, grain growth is primarily governed by the synergistic effects of three factors: grain size $D$, GB step energy $\varepsilon^*$, and the variable $n$ that is related to GSD. Therefore, it is crucial to further comprehensively investigate how these three variables jointly control grain growth behavior and to identify the underlying mechanisms responsible for the emergence and disappearance of non-Arrhenius grain growth behavior.

Based on Eq. 3, a critical interface for grain growth can be plotted within the three-dimensional variable space defined by $D$, $\varepsilon^*$, and $n$, as shown in Fig. 5. Here, a grain growth velocity less than or equal to $10^{-13}$ m/s is considered effectively zero, indicating growth stagnation. At this point, the critical interface for grain growth divides the variable space into a growth region and a stagnation region in Fig. 5A. The grains within the growth region will continue to grow, while those in the stagnation region will remain in a stagnant growth state. The critical growth line in Fig. 5A shows that the threshold value of $n$ required for grain growth in polycrystalline systems significantly decreases as the grain size decreases. Therefore, in polycrystalline systems with similar GSDs (*i.e.*, the same range of $n$ value), the smaller the grain size, the easier the $n$ values corresponding to some grains exceed the threshold value of $n$ required for their growth (especially at the nanoscale), which leads to preferential growth of some larger grains in the system. This may explain the poorer thermal stability of nanocrystalline systems and the reason why AGG is commonly observed in nanocrystalline materials. The classical theories usually attribute the poorer thermal stability of polycrystalline systems with smaller grains to their higher surface/interfacial free energy resulting from size effects. However, as highlighted in previous literature, it can be concluded that the GSD (or variable $n$) also plays a critical role in grain growth [8].

Additionally, the threshold value of $n$ required for grain growth decreases significantly with decreasing GB step energy $\varepsilon^*$, as shown in Fig. 5B-5C. The reduction in GB step energy $\varepsilon^*$ can be achieved by increasing the temperature or modifying the GB environment (*e.g.*, through doping, applying pressure, etc.) [5,8]. At lower temperatures (corresponding to higher step energy $\varepsilon^*$), the stagnation region in the variable space becomes larger, as shown in Fig. 5B, and the threshold values of $n$ required for grains growth are also higher. This means that at lower temperatures, grain growth can be suppressed by narrowing the GSD (*i.e.*, reducing the $n$ values of the grains to below their growth thresholds). As the temperature rises, the corresponding GB step energy $\varepsilon^*$ decreases, leading to a shift of the growth critical line toward smaller $n$ values in Fig. 5C. Notably, the larger the grain size, the more significant the reduction in the threshold value of $n$ required for grain growth (as illustrated in Fig. 5B). This explains why grains tend to sustain growth more easily at higher temperatures, in contrast to lower temperatures where grains tend to stagnate after some initial growth.

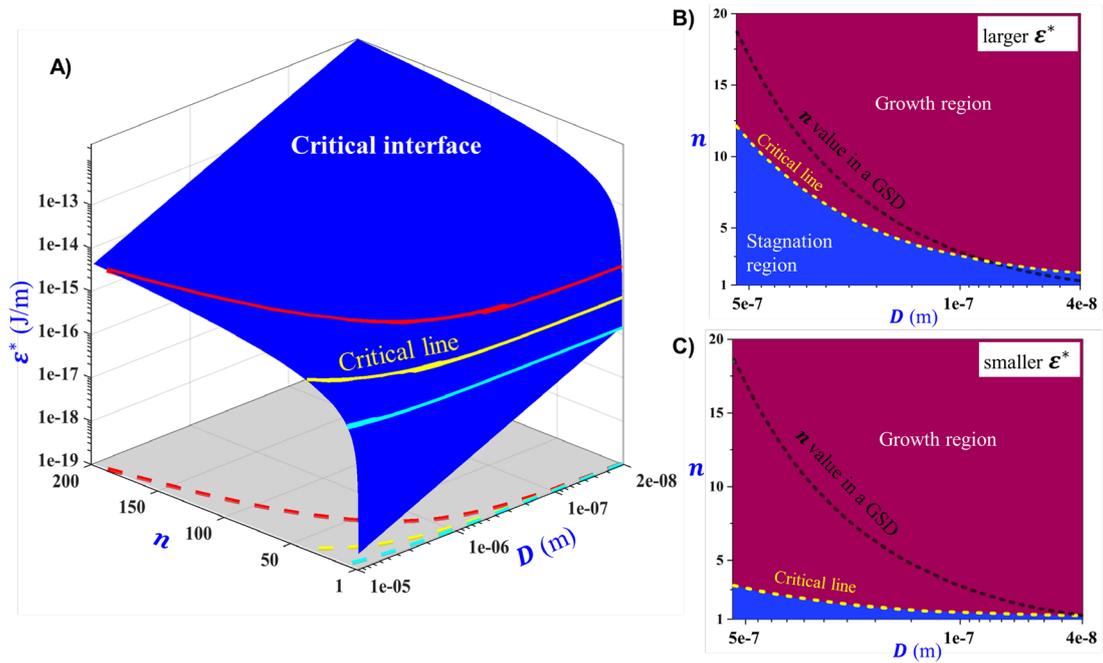

Fig. 5. **A)** Critical interface and critical lines for grain growth in the variable space defined by $n$, $D$ and $\varepsilon^*$; **B)** Projection of the critical line with larger $\varepsilon^*$, **C)** Projection of the critical line with smaller $\varepsilon^*$. The black dashed line illustrates the $n$-values associated with grains in a polycrystalline system (shown as an example).

For polycrystalline systems, changes in the threshold value of $n$ for grain growth directly influence the fraction of grains in the growth state. The higher the sintering temperature, the smaller the threshold value of $n$, and the greater the fraction of grains capable of growing. This prediction has been validated by grain growth experiments shown in Fig. 1. At lower sintering temperatures, fewer grains exhibit abnormal growth. Due to the longer incubation period, the initial grain growth velocity is slower, leading to a more gradual increase in grain size (as shown in Fig. 1A for samples sintered at 925 °C and 950 °C, as well as in Fig. 3A). As the sizes of abnormally growing grains increase, their corresponding $n$ values (the ratio relative to the matrix grains) before mutual impingement also increase, leading to an accelerated growth velocity and a rapid enlargement of grain size. In contrast, at higher sintering temperatures, the fraction of abnormally growing grains in the sample increases significantly, and the incubation period becomes shorter or entirely disappears, as shown in Fig. 1A for samples sintered at 965 °C and 975 °C as well as in Fig. 3B-3D. At lower temperatures, the incubation period lasts for tens of minutes or more, whereas at higher sintering temperatures, it is reduced to just a few minutes or even less as shown in Fig. 6. As the holding time increases or the temperature continues to rise, AGG with bimodal GSD and grain growth stagnation with monomodal distributions may alternate [8]. When the temperature exceeds GB roughening transition temperature (where the step energy $\varepsilon^*$ becomes zero), stagnating grains disappear, and the GSD becomes mono-modal.

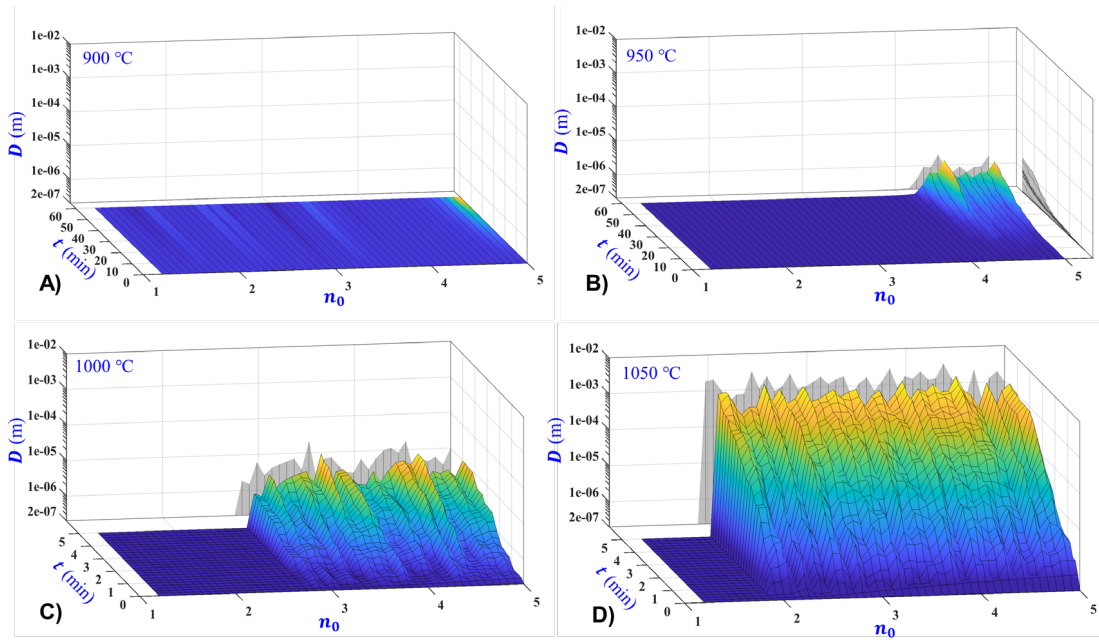

Fig. 6. Numerical simulation of the growth behavior of individual grain in a polycrystalline system at different temperatures under the same $n$-value range conditions. **A**) 900 °C, **B**) 950 °C, **C**) 1000 °C and **D**) 1050 °C. This simulation reveals how temperature (or GB step energy) influences the growth behavior of abnormally growing grains when the corresponding $n$-value decreases sharply following mutual impingement. The size of abnormally growing grains after mutual impingement is set to 200 nm.

The experimental and simulation results above demonstrate that non-Arrhenius grain growth behavior is more likely to occur in the lower-temperature range of AGG stage.

This is because, at low sintering temperatures, the growth $n$-value thresholds for grains of different sizes are relatively high, and the incubation period required for matrix grains to transition from stagnation to growth is longer. As a result, the number fraction of abnormally growing grains in the polycrystalline system remains comparatively small. When the holding times is prolonged during sintering, these abnormal growing grains accelerate in growth until mutual impingement reduces their corresponding $n$ values, leading to growth stagnation. Meanwhile, due to the longer incubation period, the stagnant matrix grains are gradually consumed during the incubation stage. Thus, only a minority of grains are able to grow while the majority of matrix grains are consumed, ultimately resulting in a polycrystalline system comprising solely these abnormally grown large grains. As the temperature increases, the fraction of abnormally growing grains increases, and the incubation period shortens. Extended holding times enable abnormally growing grains to rapidly grow by consuming the surrounding matrix grains before mutual impingements lead to growth stagnation. Meanwhile, the shortened incubation period may also allow some matrix grains to transition from a stagnant state to abnormal growth during this process. Within a certain temperature range (e.g., 975 °C and below in Fig. 2C), the grain growth processes at different sintering temperatures follow similar patterns. At higher sintering temperatures in this range, the number fraction of growing grains per unit volume is relatively larger, while the corresponding matrix grains available for consumption are fewer. Once the matrix grains are consumed, the growing grains enter a stagnation state due to mutual impingement. Consequently, samples sintered at higher temperatures, which contain more growing grains, ultimately develop smaller final grain sizes. This explains the counterintuitive phenomenon where samples sintered at lower temperatures exhibit larger grain sizes, leading to the so-called non-Arrhenius grain growth behavior. Furthermore, as shown in Fig. 2C for samples at temperatures below 965 °C, the lower the sintering temperature, the larger the grain size, confirming these predictions. Thus, the non-Arrhenius behavior of grain growth always occurs during the AGG stage. This behavior is influenced not only by temperature and temperature-dependent variables (such as GB step energy $\varepsilon^*$) but also by the GSD-related variables $n$, which plays a crucial role.

As the sintering temperature continues to increase (e.g., above 975 °C in Fig. 2C), the continuous reduction of $\varepsilon^*$ leads to a lower threshold value of $n$ for growth and a significantly shortened incubation period. Some abnormally growing grains that impinge with on one another may also continue to grow due to lower threshold value of $n$, as shown in Fig. 6B-6D. At this stage, grain size continues to increase, leading to the disappearance of the previously observed non-Arrhenius grain growth phenomenon and a reversion to Arrhenius-type behavior, as shown in Fig. 2C for samples sintered at 975 °C and above. This transition in grain growth behavior, from non-Arrhenius to Arrhenius with increasing temperature, is experimentally verified in Fig. 2C and is consistent with previously reported experimental findings [14]. Furthermore, during the AGG stage characterized by non-zero step energy, grain growth may exhibit alternating growth and stagnation states as temperature continues to rise [8,19], where non-Arrhenius grain growth behavior may also recur. It can therefore be inferred that the occurrence of non-Arrhenius behavior is not confined to a specific temperature range. In polycrystalline systems with larger grain sizes, the temperature at which non-Arrhenius behavior occurs also increases accordingly, as reported in the literature at 1350 °C for $SrTiO_3$ ceramics [10]. Thus, the occurrence of non-Arrhenius grain growth behavior is influenced not only by temperature but also by the grain size and its distribution in the polycrystalline system. This phenomenon does not stem from a reduced GB migration velocity but is instead associated with the fraction of abnormally growing grains and the duration of the incubation period of matrix grains.

## 4. Conclusion

The emergence and disappearance of non-Arrhenius grain growth behavior can be effectively explained by the general grain growth model. The behavior of non-Arrhenius grain growth occurs with a thermally activated GB mobility during the lower-temperature stage of AGG, wherein the lower the temperature, the smaller the proportion of grains activated for growth and the longer incubation period for grain growth. These activated grains then consume a larger fraction of the stagnant matrix grains at lower temperature, thereby achieving a larger final grain size. The fraction of abnormally growing grains within polycrystalline systems is synergistically controlled by temperature-dependent parameters (such as GB step energy and GB energy) and temperature-independent factors (such as grain size and its distribution). When temperature-independent factors are held constant, this fraction increases with increasing temperature or decreasing GB step energy, which influences the occurrence of non-Arrhenius behavior and the subsequent emergence of Arrhenius-type behavior at elevated temperature. Moreover, the coexistence of distinct GB types (or mobilities) is not required for the occurrence of non-Arrhenius grain growth behavior. Consequently, the occurrence of non-Arrhenius grain growth does not exhibit a definitive characteristic temperature due to the effects of temperature-independent parameters.

## CRediT authorship contribution statement

**Xinlei Pan:** Investigation, Visualization, Software, Validation, Formal analysis, Data curation. **Jingyu Li：** Visualization, Data curation. **Jianfeng Hu:** Conceptualization, Writing – original draft, Writing – review & editing, Methodology, Investigation, Formal analysis, Data curation, Funding acquisition.

## Declaration of competing interest

The authors declare that they have no known competing financial interests or personal relationships that could have appeared to influence the work reported in this paper.

## Acknowledgments

This work was supported by the National Natural Science Foundation of China (NSFC) under Grant No. 52173223 to J.H.

## Reference:

[1] Humphreys J, Rohrer GS, Rollett A. Grain Growth Following Recrystallization. In: Recrystallization and Related Annealing Phenomena, Elsevier; 2017: 375–429.

[2] Hillert M. On the theory of normal and abnormal grain growth. *Acta Metall* 1965, **13**: 227–38.

[3] MacPherson RD, Srolovitz DJ. The von Neumann relation generalized to coarsening of three-dimensional microstructures. *Nature* 2007, **446**: 1053–5.

[4] Gottstein GG, Shvindlerman LS. Grain Boundary Migration in Metals. 2nd ed. CRC Press; 2010.

[5] Hu J, Wang X, Zhang J, *et al.* A general mechanism of grain growth —I. Theory. *J Mater* 2021, **7**: 1007–13.

[6] Atkinson H V. Overview no. 65. Theories of normal grain growth in pure single phase systems. *Acta Metall* 1988, **36**: 469–91.

[7] Mullins WW. The statistical self-similarity hypothesis in grain growth and particle coarsening. *J Appl Phys* 1986, **59**: 1341–9.

[8] Hu J, Zhang J, Wang X, *et al.* A general mechanism of grain growth-II: Experimental. *J Mater* 2021, **7**: 1014–21.

[9] Bäurer M, Shih SJ, Bishop C, *et al.* Abnormal grain growth in undoped strontium and barium titanate. *Acta Mater* 2010, **58**: 290–300.

[10] Rheinheimer W, Schoof E, Selzer M, *et al.* Non-Arrhenius grain growth in strontium titanate: Quantification of bimodal grain growth. *Acta Mater* 2019, **174**: 105–15.

[11] Pan Y, Huang L, Feng J, *et al.* Non-Arrhenius behavior of grain growth in (K, Na)NbO3-based ceramics and its effect on piezoelectric properties. *Ceram Int* 2022, **48**: 37085–94.

[12] Zhang K, Weertman JR, Eastman JA. Rapid stress-driven grain coarsening in nanocrystalline Cu at ambient and cryogenic temperatures. *Appl Phys Lett* 2005, **87**: 6–9.

[13] Kelly MN, Rheinheimer W, Hoffmann MJ, *et al.* Anti-thermal grain growth in SrTiO3: Coupled reduction of the grain boundary energy and grain growth rate constant. *Acta Mater* 2018, **149**: 11–8.

[14] Zahler MP, Jennings D, Guillon O, *et al.* Non-Arrhenius grain growth in SrTiO3: Impact on grain boundary conductivity and segregation. *Acta Mater* 2025, **283**: 120560.

[15] Rheinheimer W, Hoffmann MJ. Non-Arrhenius behavior of grain growth in strontium titanate: New evidence for a structural transition of grain boundaries. *Scr Mater* 2015, **101**: 68–71.

[16] Sternlicht H, Rheinheimer W, Hoffmann MJ, *et al.* The mechanism of grain boundary motion in SrTiO3. *J Mater Sci* 2015, **51**: 467–75.

[17] Homer ER, Johnson OK, Britton D, *et al.* A classical equation that accounts for observations of non-Arrhenius and cryogenic grain boundary migration. *Npj Comput Mater* 2022, **8**: 157.

[18] Verma A, Johnson OK, Thompson GB, *et al.* Insights into factors that affect non-Arrhenius migration of a simulated incoherent Σ3 grain boundary. *Acta Mater* 2023, **258**: 119210.

[19] Jung SH, Kang SJL. Repetitive grain growth behavior with increasing temperature and grain boundary roughening in a model nickel system. *Acta Mater* 2014, **69**: 283–91.